\documentclass[NETN]{stjour}
\articletype{Research}
\usepackage{rotating}
\usepackage{tabularx}
\usepackage{booktabs}
\usepackage[english]{babel}
\usepackage{varioref}
\usepackage{hyperref}
\usepackage{cleveref}
\usepackage{changepage}
\usepackage{csquotes,booktabs,dcolumn}
\usepackage[title]{appendix}
\usepackage{textpos}
\usepackage{graphicx}

\newcolumntype{d}[1]{D{.}{.}{#1}}
\newcolumntype{Y}{>{\centering\arraybackslash}X}

\begin{document}
\title[]{A paper's corresponding affiliation and first affiliation are consistent at the country level in Web of Science}

\author[Author Names]
{Jianfei Yu\affil{1},Chunxiao Yin\affil{1*}, Linlin Liu\affil{1}, 
\and Tao Jia\affil{1*}}

\affiliation{1}{College of Computer and Information Science, Southwest University, Chongqing,400715,  P. R. China.}

\correspondingauthor{Chunxiao Yin}{yincx@swu.edu.cn}
\correspondingauthor{Tao Jia}{tjia@swu.edu.cn}

\keywords{first affiliation; corresponding affiliation; web of science; straight counting}


\begin{abstract}
\textbf{Purpose:} The purpose of this study is to explore the relationship between the first affiliation and the corresponding affiliation at the different levels via the scientometric analysis.

\textbf{Design/methodology/approach:} We select over 18 million papers in the core collection database of Web of Science (WoS) published from 2000 to 2015, and measure the percentage of match between the first and the corresponding affiliation at the country and institution level.

\textbf{Finding:} We find that a paper's the first affiliation and the corresponding affiliation are highly consistent at the country level, with over 98\% of the match on average. However, the match at the institution level is much lower, which varies significantly with time and country. Hence, for studies at the country level, using the first and corresponding affiliations are almost the same. But we may need to take more cautions to select affiliation when the institution is the focus of the investigation. In the meanwhile, we find some evidence that the recorded corresponding information in the WoS database has undergone some changes since 2013, which sheds light on future studies on the comparison of different databases or the affiliation accuracy of WoS. 

\textbf{Research limitations:} Our finding relies on the records of WoS, which may not be entirely accurate.

\textbf{Practical implications:} Given the scale of the analysis, our findings can serve as a useful reference for further studies when country allocation or institute allocation is needed.

\textbf{Originality/value:} Existing studies on comparisons of straight counting methods usually cover a limited number of papers, a particular research field or a limited range of time. More importantly, using the number counted can not sufficiently tell if the corresponding and first affiliation are similar. This paper uses a metric similar to Jaccard similarity to measure the percentage of the match and performs a comprehensive analysis based on a large-scale bibliometric database.
\end{abstract}


\section{Introduction}

When investigating the scientific productivity, scientific impact or scientific development of a country, we need to first allocate papers to different countries \citep{gazni2012mapping, gingras2018assessing, gonzalez2016scientific}. This problem is partially related to how we ``count'' papers, giving rise to a rich body of studies on the counting method \citep{waltman2015field, aksnes2012ranking,huang2011counting, gauffriau2008comparisons, vavryvcuk2018fair, korytkowski2019publication, smolinsky2020co}. Such studies become more important given the intensified international collaboration nowadays, through which the scientific production is characterized by not only multiple institutes but also by multiple countries \citep{zacharewicz2019performance, gul2015middle}. Among the counting methods commonly applied, straight counting is the one that allocates the whole credit of a paper to a single entity \citep{gauffriau2008comparisons,lin2013the, huang2011counting}. In other words, the paper would belong to one single country or one institute among the multiple affiliations of the paper. Previous studies suggest that straight counting is preferred in professional and scientific bibliometrics operations, especially when dealing with large-scale literature data \citep{huang2011counting, larsen2008state}.

The logic behind the straight counting is that one of the most prominent affiliations (or authors) owns the whole paper \citep{hagen2014counting, mattsson2011correspondence}. Existing studies mainly consider two options: using the first affiliation (author) \citep{van2009strength, gauffriau2007publication, borner2006mapping} or the corresponding affiliation (author) \citep{man2004some, mazloumian2013global}. The counting results by the two options are compared. In some previous studies, it is found that counting results using the first affiliation and the corresponding affiliation are consistent in reflecting research productivity at the country level \citep{huang2011counting, waltman2015field}. Nevertheless, these studies usually focus on a small fragment of paper data, covering only a limited research field or time range, which may not be conclusively generalized to other circumstances. 

More importantly, country allocation is related to applications more general than just counting \citep{petersen2019methods}. For instance, when studying the citation network of countries, we need to assign a country to each paper \citep{radicchi2009diffusion, apolloni2013collaboration, bornmann2014ranking, hu2020global}. As a paper would cite multiple papers and be cited by other papers, the country attribute of a node in the citation network will affect nodes connected to this node, which eventually affects the topological representation of the network. Likewise, in recent emerging interdisciplinary studies of ``science of science'' \citep{fortunato2018science, zeng2017science}, the patterns uncovered rely on a variety of features, such as a paper's team composition, its financial supports, and its research topics \citep{liu2020china, wu2019large, guo2019contributions, shen2014collective, milojevic2014principles, jia2017quantifying, thelwall2019gender, chen2020rank, selten2020longitudinal}. What a paper meant in these studies is more than a number. Hence, we argue that existing results on the counting method are insufficient for the problem of country allocation. As an example, let us consider a case that half of the papers in our data have country A in the first affiliation and country B in the corresponding affiliation, whereas for the other half of the papers the first affiliation is associated with country B and the corresponding affiliation is associated with country A. In terms of the counted number of papers, using either the first or the corresponding affiliation are the same. But the country allocation by the first affiliation is entirely different from that by the corresponding affiliation.

In this study, we perform a comprehensive analysis using a large bibliometric database that contains over 18 million papers published from 2000 to 2015 on the Web of Science (WoS). Instead of counting, we measure the percentage of matches between the first affiliation and the corresponding affiliation. We find that the two affiliations of a paper are consistent at the country level, with over 98\% of matches on average. Therefore, for studies at the country level, either counting the number of papers or the construction of citation networks, results based on the first or corresponding affiliation are almost the same. The match at the institution level is lower, which also varies significantly with time and country. Hence we may need to take more cautions to select affiliation when the institution is the focus of the investigation. Given the large scale of the analysis, our results can serve as a useful reference for further research when the country allocation or institute allocation is needed. The analyses also reveal the existence of record changes in the WoS database, which bring varaitions on how the corresponding affiliation is recorded. This finding may shed lights on future studies on the comparison of different database or the affiliation accuracy of WoS data. 

\section{Data and Method}
{\bf Data set.} We use the data in the Web of Science (WoS), which is a well-established database used for the bibliometric analysis \citep{han2014international}. The data covers the Science Citation Index Expanded (SCIE) database, the Social Sciences Citation Index (SSCI) database, and the Arts \& Humanities Citation Index (A \& HCI) database. In total, we analyze over 18 million papers published from 2000 to 2015, which include articles, notes, reviews, letters, and conference proceeding papers.

{\bf Countries considered.} The most productive 16 countries, in terms of the number of scientific papers, are selected for our study. They are the United States (US), China (CN),the United Kingdom (GB), Germany (DE), Japan (JP), Italy (IT), France (FR), Canada (CA), India (IN), Korea (KR), Spain (ES), Australia (AU), Brazil (BR), Netherlands (NL), Turkey (TR), and Russia (RU). The total number of papers by these 16 countries is 18,432,794, covering over 76\% of worldwide papers. 

{\bf Address information.} WoS records the list of affiliations and the order of these affiliations for each paper. Starting in 2008, WoS also records the list of affiliations of each author. WoS specifically records the reprint affiliation of each paper, which is considered to be equivalent to the corresponding affiliation \citep{kahn2016return, fox2018patterns, wang2013international, gonzalez2017dominance, duffy2017last}. In some very recent records, one paper may have multiple reprint affiliations. This is, however, not commonly observed in papers published during the period analyzed in this work. Almost all papers have only one reprint affiliation.

{\bf Comparing Institution Names.} The institution names in the meta data are usually not consistent \citep{donner2020comparing, rimmert2017disambiguation}. One institution can be referred by different ways related to different writing and coding rules (i.e., by the official, full institution name, and/or by varying forms of abbreviations). For example, “Tsinghua University” and “Tsinghua Univ” are the same institution, where the former is the full name and the latter is the abbreviated form. Likewise, two names with similar word composition may be related to two distinct institutions. For example, “Univ Colorado” and“Univ Colorado Denver” are different, with the latter being a branch of the former. In general, the name comparison is related to a broader and more challenging problem called institution name disambiguation \citep{huang2014institution, jacob2014scool}.

Fortunately, in this work, we only need to determine if two institutions are the same in one single paper. It is unlikely that a paper contains two distinct yet literally similar institutions as its reprint and first affiliation. Therefore, we do not need to solve the name disambiguation problem. Here, we measure the edit distance of two institutions in the first and reprint affiliation . The edit distance, also called Levenshtein distance, is defined as the minimum number of edits needed to transform one string into the other \citep{levenshtein1966binary}. We set the threshold as 90\%. Two institution names with edit similarity equal or greater than this threshold are considered as the same name.

{\bf Calculating the Match.} We use a metric similar to Jaccard similarity to measure the percentage of the match between the first and corresponding affiliation. In particular, we have 
\begin{align}
{P_i} = \frac{|C_i \cap F_i |}{|C_i|},
\end{align}
where $C_i$ is the set of papers whose the corresponding affiliation is associated with $i$ (which can be a country or a institution) and $F_i$ is the set of papers whose the first affiliation is associated with $i$. We also test the results by changing the denominator to $|F_i|$ and $|C_i \cup F_i |$. The conclusion does not change with such variations.

\section{Results}
\subsection{The match between the first affiliation and the corresponding affiliation at the country level}
We first compare the country in the first and the corresponding affiliation of a paper. The statistics demonstrate a high consistency at the country level (Figure 1a). In 98.57\% of all papers analyzed, the first and the corresponding affiliation point to the same country. China ranks the first among the 16 countries, with the percentage of the match $P_\text{ctry} = 99.58\%$ and Canada ranks the last with $P_\text{ctry} =97.33\%$.

We further analyze how the match changes over time. In general, $P_\text{ctry}$ of all countries are high at different years. Although there is a sharp decline of $P_\text{ctry}$ occurring in the year 2012, the lowest value is above 94\% (Figure 1b). In general, we can conclude that the country in the first and the corresponding affiliation of a paper have a high percentage of mach for different countries and in different years.

It is noteworthy that the label of the corresponding affiliation in the WoS may not be very accurate \citep{huang2011counting, moya2013research}. Indeed, for some countries we find a high percentage of papers (such as 76.34\% in China and 72.8\% in India) whose corresponding affiliation is also the first affiliation. This gives rise to a concern on validity of our conclusion as the high percentage of match can be simply a result of high overlap between the first and the corresponding affiliation. For this reason, we focus on papers whose the corresponding and the first affiliation are different. The percentage of match decreases slightly, with $P_\text{ctry} = 97.52\%$ on average (Figure 1c). Russia becomes the country with the highest match. The $P_\text{ctry}$ in different years also remain at a high level, with the lowest value 90\%. In other words, even when the first and corresponding affiliation are different, we still have at least 90\% of match at the country level. The result supports our conclusion that the first and the corresponding affiliation are highly consistent at the country level.

Finally, both Figure 1b and Figure 1d show a sharp decline of $P_\text{ctry}$ in the year 2013 which virtually splits the curve into two phases (2000-2012 and 2013-2015). $P_\text{ctry}$ in the first phase is higher (on average 98.96\% in Figure 1b and 98.15\% in Figure 1d) than that in the second phase (on average 96.90\% in Figure 1b and 94.82\% in Figure 1d). Something happens in 2013 that brings down the overall percentage of the match by roughly 3 percentage points. Although the overall consistency is high and not significantly affected by this decline, this phenomenon needs further exploration which will be discussed in detail later. 

\begin{figure}[htp]
\centering
\includegraphics[width=1.00\textwidth]{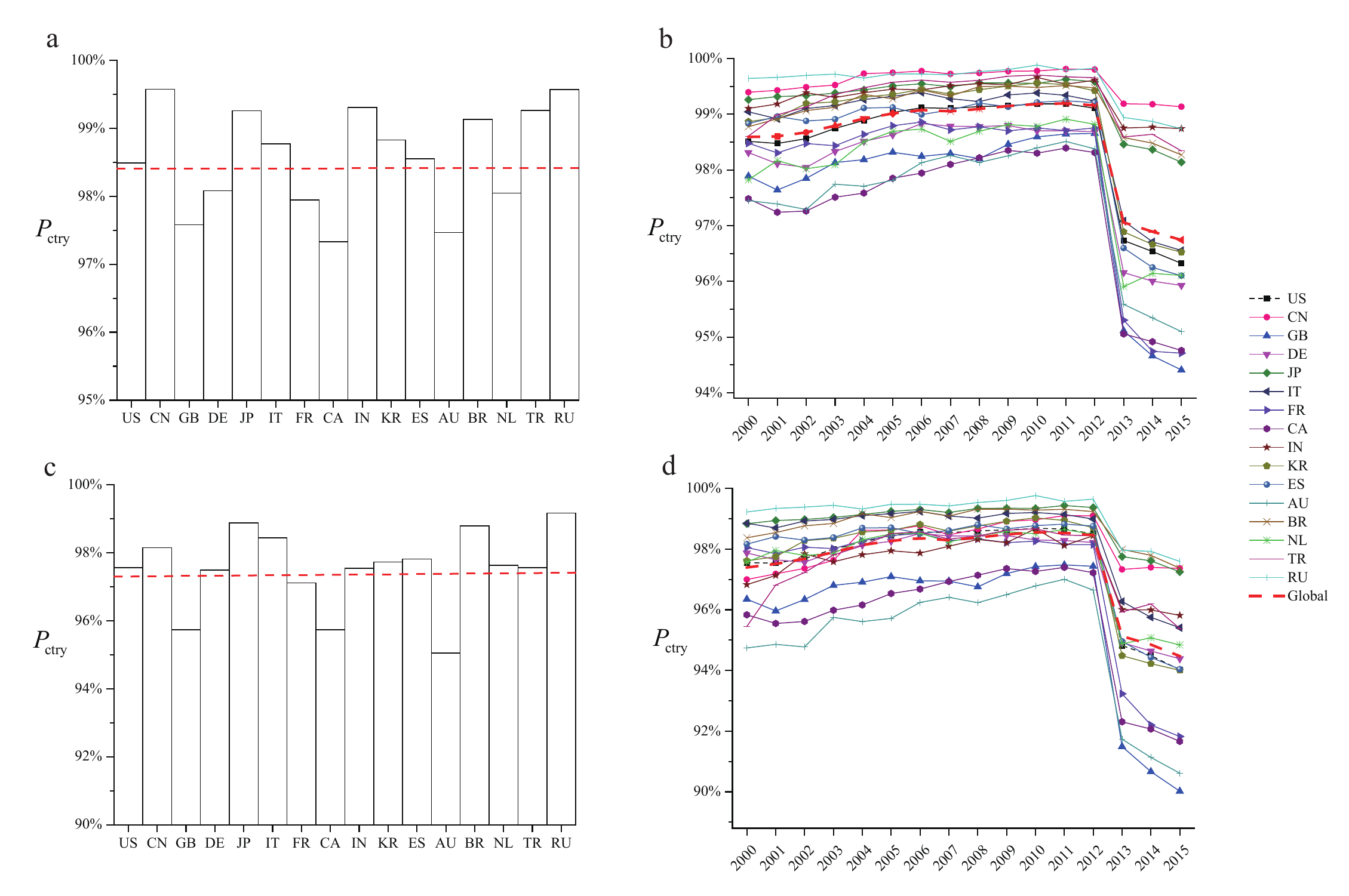}
\caption{{\bf (a)} The percentage of affiliation match at the country level $P_\text{ctry}$ for all papers published during 2000 to 2015. The dashed line corresponds to the global average. {\bf (b)} $P_\text{ctry}$ for papers published at different years also remain at a high level. {\bf (c)} The percentage of affiliation match at the country level $P_\text{ctry}$ for all papers whose first and corresponding affiliation are different. The dashed line corresponds to the global average. {\bf (d)} The chang of $P_\text{ctry}$ over time for papers published at a given year whose first and corresponding affiliation are different.}
\label{fig:fig1}
\end{figure}

\subsection{Match between the first affiliation and the corresponding affiliation at the institution level}
Since there are thousands of institutions all over the world, it is impossible to show the results for each institution. Therefore, we use the average value grouped by countries of their institutions. As shown above, the country information in the first and corresponding affiliation are highly consistent, using either of them should give roughly the same results. In particular, let $P_\text{inst}$ denote the percentage of papers with the first and the corresponding affiliation matched at the institution level in each country. We find that $P_\text{inst}$ is lower than $P_\text{ctry}$, whose global average is 91.43\% (Figure 2a). The $P_\text{inst}$ also demonstrates a large variety among different countries. China is the highest with $P_\text{inst}$ = 97.23\% and the $P_\text{inst}$ of Brazil is the lowest (86.63\%), with a difference about 10\% (Figure 2a). 

\begin{figure}[htp]
\centering
\includegraphics[width=1.00\textwidth]{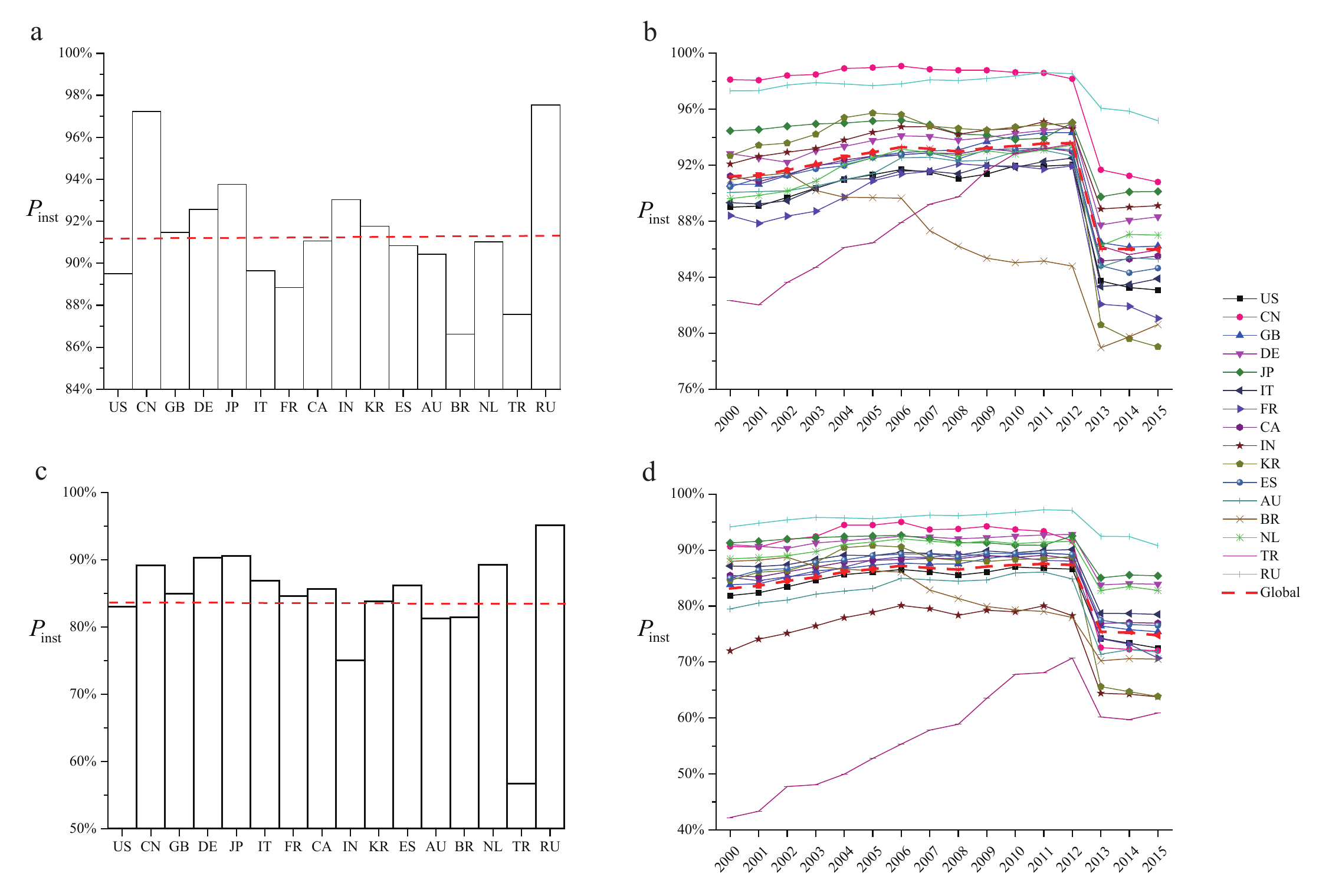}
\caption{{\bf (a)} The percentage of affiliation match at the institution level $P_\text{inst}$ for all papers published during 2000 to 2015. The dashed line corresponds to the global average.  {\bf (b)} $P_\text{inst}$ for papers published at different years, which is lower than $P_\text{ctry}$. {\bf (c)} The percentage of affiliation match at the institution level $P_\text{inst}$ for all papers whose first and corresponding affiliation are different. The dashed line corresponds to the global average. {\bf (d)} The change of $P_\text{inst}$ over time for papers published at a given year whose first and corresponding affiliation are different.}
\label{fig:fig2}
\end{figure}

There is also a large change of $P_\text{inst}$ over time. Especially, except for China and Russia, the $P_\text{inst}$ of other countries are between 88\% to 94\%, and after 2012 the $P_\text{inst}$ of most countries is below 90\%. For Brazil and Korea, this value is even below 80\%. The match at the institution level is not high in all countries at all years. Hence, allocation by the institution in the first affiliation may give different results compared with that by the institution in the corresponding affiliation. One may need to carefully exam the robustness of the conclusion if the analysis is at the institution level.

Similar to the analysis at the country level, we also conduct an analysis by excluding papers whose the first affiliation is labeled as the corresponding affiliation. The results in Figure 2c indicate that the $P_\text{inst}$ is much lower, with 84.02\% on average. And the highest $P_\text{inst}$ in most countries located between 80\% to 90\% (Figure 2d). In Turkey, as an extreme case, the lowest $P_\text{inst}$ is only 42.2\% in 2000. This further supports our conclusion that the first and corresponding affiliation are not consistent at the institution level.

It is noteworthy that the sharp decline from 2012 to 2013 observed at the country level is also found at the institution level. The drop is even higher. Specifically, there is a 7 percentage points drop in Figure 2b on average and 12 percentage points drop in Figure 2d. This further urges us to explore the cause of the sharp decline.

\subsection{The sharp decline caused by the record change in WoS}

\begin{figure}[H]
\centering
\includegraphics[width=1.0\textwidth]{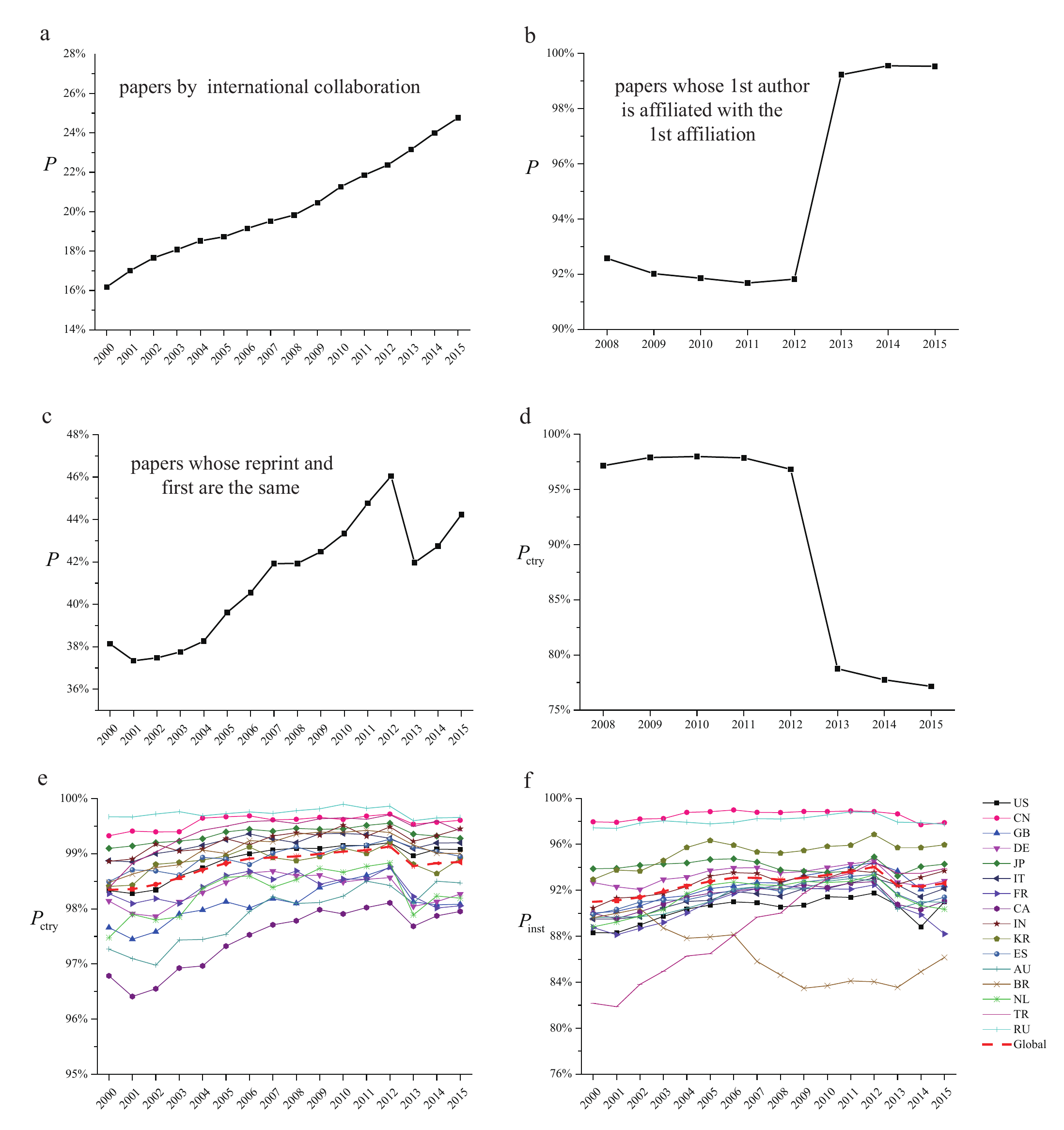}
\caption{{\bf (a)} The percentage of papers produced by international collaboration. The increase is gradual and steady with no sudden or significant changes. {\bf (b)} The percentage of papers whose first affiliation is included in the affiliation list of the first author. Before 2013, about 8\% of papers have the first author not affiliated with the first affiliation. {\bf (c)} The percentage of papers whose reprint and first affiliation are the same. The value has a sharp decrease in 2013. {\bf (d)} We consider papers whose first author is also the corresponding author and measure the percentage of affiliation match $P_\text{ctry}$ at the country level. $P_\text{ctry}$ has a sharp decrease in 2013. {\bf (e, f)} The percentage of affiliation match at the country level $P_\text{ctry}$ and at the insitution level for $P_\text{inst}$ papers whose first author serves as the corresponding author. The statistics no longer demonstrate the sharp decline.}
\label{fig:fig3}
\end{figure}

There are two reasons that seem capable of explaining the sharp decline. Since the decline is observed at both the country and institution level, one possibility is that the number of internationally co-authored papers has an acceleration in 2012, which gives rise to a sudden decrease in the percentage of the matched countries. The other possibility is that the decline is simply caused by the manner that WoS records the data. In other words, the label position of the corresponding affiliation has changed in the WoS since 2012.

To test the first hypothesis, we calculate the percentage of papers in our data that are produced by international collaboration. The international collaborated papers are those that contain affiliations in different countries \citep{gazni2011investigating, iwami2020current}. We observe that the output by international collaboration demonstrates an increasing trend (Figure 3a), which is in line with that in previous studies \citep{gazni2012mapping,lariviere2015team}. Nevertheless, the increase is gradual and steady. There is no sudden or significant change on the extend of international collaboration. Therefore, the decline observed in Figures 1 and 2 can not be attributed to patterns of international collaboration.

For the second possibility, we indeed notice certain changes in the WoS data occurring in 2013 that can generate some drastic fluctuations in the statistics. For example, WoS records two types of address information. One is the affiliation list of the paper and the other is the affiliation list of each author. Ideally, the affiliation list of the first author should contain the first affiliation of the paper \citep{nederhof1993modeling, larsen2008state}. But before the year 2013, there are roughly 8\% of papers whose first affiliation is not included in the affiliation list of the first author (Figure 3b). The turning point appears in 2013. Since then the first author is almost always affiliated with the first affiliation (Figure 3b). Likewise, the fraction of papers whose reprint and first affiliation are the same increases with time. But the value has a sharp decrease in 2013 (Figure 3c). These sudden changes imply some updates in WoS data set. However, the patterns observed in Figures 3b and 3c can not explain the sharp decrease observed in Figures 1 and 2. When we remove papers whose first author is not affiliated with the first affiliation, the sharp decline still preserves. In Figures 1b and 2b, we have already shown that $P_\text{ctry}$ and $P_\text{inst}$ suddenly decreases in 2013 when removing papers whose first and corresponding affiliation are the same. 

What we find most relevant to the sharp decrease is the change in the records of the corresponding author. In the WoS data, the percentage of papers whose first author does not serve as the corresponding author increases smoothly with time. But if we focus on these kinds of papers, we can observe a sudden decrease in the percentage of papers whose first and corresponding affiliation are the same (Figure 3d). If we remove these papers in our analysis and consider only papers whose first author is also the corresponding author, the sharp decline in $P_\text{ctry}$ and $P_\text{inst}$ are no longer observed (Figures 3e and 3f). Therefore, we believe that it is the change of the corresponding author records that gives rise to the sudden drop of the matched affiliation at the country and the institution level.

\section{Conclusion}
To summarize, we analyze over 18 million papers in the WoS database published from 2000 to 2015. We find that a paper’s the first affiliation and the corresponding affiliation are highly consistent at the country level, with over 98\% of the match on average. The extend of the match varies slightly when we focus on different years or consider only the circumstance when the first and the corresponding affiliation are different. Nevertheless, the match remains at a high level, with the lowest over 90\%. The result is in line with previous findings that straight counting by the first and the corresponding affiliation give rise to close numbers. But our result can be applied to more general applications. When allocating a country to a paper, using the first or the corresponding affiliation would yield roughly the same results. Considering the fact that the corresponding affiliation is not usually explicitly given (in Microsoft Academic Graph for example \citep{wang2020microsoft, ranjbar2018accuracy}), our finding can be a useful reference for future studies that require country allocation. 

We also find that the mach at the institution level is much lower. On average, about 10\% of the time, one would get different results when allocating the institution by the first affiliation instead of the corresponding affiliation. The difference may not be significant when only the number of papers is concerned. But for extended studies such as the impact, the research behavior and the collaboration pattern of different institutions, we need to be more cautious in deciding which institution a paper belongs to. At least, the robustness of the conclusion needs to be tested by different allocation methods. This also raises interesting questions on the university ranking \citep{lin2013the, chen2020rank, abramo2015evaluating, selten2020longitudinal}, whose results rely on how the scientific output by different universities are grouped.

Finally, we observe some drastic changes in WoS records that bring a sharp decline in our measures. In particular, the change of corresponding author records gives rise to a lower match at the country and institution level. There are studies analyzing and comparing different data sets of publications \citep{adriaanse2013web, lopez2008coverage, aghaei2013comparison, falagas2008comparison}. Some studies also question the accuracy of WoS data in citations and topic classifications \citep{franceschini2016empirical, ranjbar2018accuracy, van2019accuracy}. Except for a few works, however, the accuracy of the affiliation information is not well discussed. Our observation implies that the affiliation of a certain fraction of papers may not be accurately recorded in WoS before 2013. 
At least, some papers in the WoS may not have the correct corresponding information. The decrease in the statistics also implies that records after 2013 may have a better accuracy than before.
The potential errors in WoS data naturally raise concerns about the validity of our findings. If there are flaws in the data we analyzed, to what extend could we generalize the conclusion that a paper's corresponding and first affiliation are consistent at the country level. Note that, however, some data sets may have better accuracy at some certain records, but none of them are perfect. If we inevitably need to utilize the imperfect data to perform extended and comprehensive research, we need to tolerate certain errors within it. From that perspective, we believe that our finding is still useful, at least for those research relying on WoS data. Based on what WoS tells, the percentage of the match $P_\text{ctry}$ is very high in different periods of time and different sets of papers considered. Our finding also provides a reference point if other data sets are considered. Given the size of the data analyzed, it is hard to manually check the accuracy of the affiliation records of WoS. It would be meaningful and interesting to find an automatic approach to perform a large-scale exam on the corresponding affiliation and author records in WoS data.

\newpage

\bibliography{ref}


\acknowledgments
We thank Prof. Barabasi at CCNR for giving access to the WOS data.

\authorcontributions 
T.J. designed the research, T.J. did the data parsing and cleaning, T.J., J.F. and L.L analyzed the data, collected the statistics and reviewed related literature. T.J., C.X. and J.F. prepared the initial draft of the manuscript. All authors contributed comments on the results and revisions to the final version.

\end{document}